\documentclass{article}
\usepackage{dcase2021,amsmath,graphicx,url,times,booktabs,tabularx,amsfonts,siunitx,microtype}
\usepackage[numbers]{natbib}

\title{An Encoder-Decoder Based Audio Captioning System With Transfer and Reinforcement Learning}

\name{
      Xinhao Mei$^{1}$,
      Qiushi Huang$^{1}$,
      Xubo Liu$^{1}$, 
      Gengyun Chen$^{2}$,
      Jingqian Wu$^{3}$\sthanks{Jingqian Wu is currently with Wake Forest University, USA},
      Yusong Wu$^{3}$\sthanks{Yusong Wu is currently with University of Montreal, Canada},
      }
\secondlinename{
      Jinzheng Zhao$^{1}$,
      Shengchen Li$^{3}$,
      Tom Ko$^{4}$, 
      H Lilian Tang$^{1}$, 
      Xi Shao$^{2}$,
      Mark D. Plumbley$^{1}$, 
      Wenwu Wang$^{1}$ 
      }
\address{$^1$ University of Surrey, Guildford, United Kingdom\\ 
        $^2$ Nanjing University of Posts and Telecommunications, Nanjing, China \\
        $^3$ Xi'an Jiaotong-Liverpool University, Suzhou, China \\
        $^4$ Southern University of Science and Technology, Shenzhen, China\\
 }

\begin{document}

\ninept
\maketitle

\begin{sloppy}

\begin{abstract}
Automated audio captioning aims to use natural language to describe the content of audio data. This paper presents an audio captioning system with an encoder-decoder architecture, where the decoder predicts words based on audio features extracted by the encoder. To improve the proposed system, transfer learning from either an upstream audio-related task or a large in-domain dataset is introduced to mitigate the problem induced by data scarcity. Besides, evaluation metrics are incorporated into the optimization of the model with reinforcement learning, which helps address the problem of ``exposure bias'' induced by ``teacher forcing'' training strategy and the mismatch between the evaluation metrics and the loss function. The resulting system was ranked 3rd in DCASE 2021 Task 6. Ablation studies are carried out to investigate how much each element in the proposed system can contribute to final performance. The results show that the proposed techniques significantly improve the scores of the evaluation metrics, however, reinforcement learning may impact adversely on the quality of the generated captions.


\end{abstract}


\section{Introduction}
\label{sec:intro}

An automated audio captioning (AAC) system describes an audio signal using natural language, which is a cross-modal translation task involving the technologies of audio processing and natural language processing \cite{drossos2017automated}. Generating a meaningful description of an audio clip not only requires recognizing what audio events are presented, but also their properties, activities as well as spatial-temporal relationships. Audio captioning could be useful in several applications, such as subtitling for sound in a television program, assisting the hearing-impaired to understand environmental sounds, and analysing sounds in smart cities for security surveillance.

\citet{drossos2017automated} proposed the initial work in audio captioning, where they introduced an encoder-decoder architecture based on recurrent neural networks (RNNs) on a commercial sound effects library, ProSound Effects. After that, with the release of two freely available datasets AudioCaps \cite{kim2019audiocaps} and Clotho \cite{drossos2020clotho}, and a new audio captioning task in DCASE challenges, this field has received increasing attention. Almost all researchers investigating audio captioning have utilised an encoder-decoder architecture based on deep neural networks (DNNs). For the encoder, recurrent neural networks (RNNs) \cite{drossos2017automated, kim2019audiocaps, nguyen2020temporal}, convolutional neural networks (CNNs) \cite{chen2020audio, tran2020wavetransformer}, or their combinations, i.e. convolutional recurrent neural network (CRNN) \cite{xu2020crnn}, have been used to model the temporal, or temporal-spectral relationship between audio features. For the decoder, recurrent neural network (RNN) has been widely used to generate captions by decoding audio features to text descriptions \cite{drossos2017automated, kim2019audiocaps, nguyen2020temporal,xu2020crnn}. To align the cross-modal representation between audio and language, attention mechanisms with different implementation methods have been used between the encoder and decoder \cite{kim2019audiocaps, wang2020automated}. With the popularity of Transformer in natural language processing (NLP) and computer vision (CV), some researchers try to use Transformer as the decoder \cite{chen2020audio,koizumi2020transformer,takeuchi2020effects}. In addition, keywords and semantic information predicted from the input audio are introduced to guide caption generation \cite{nguyen2020temporal, koizumi2020transformer}. The encoder-decoder architecture with ``CNN-Transformer'' was shown to offer better performance in the DCASE 2020 challenge \cite{chen2020audio}, which is chosen as the baseline system in our work. 

As the availability of data in the audio captioning task is limited, training an end-to-end cross-modal audio captioning system from scratch becomes even more difficult. Transfer learning has been widely used to solve this data scarcity problem, where pre-trained audio models from an upstream audio processing task (i.e. audio tagging and sound event detection) are utilized as the audio encoder \cite{chen2020audio, xu2021investigating}. As using pre-trained audio models can only transfer knowledge in the audio modality, we also pre-train the whole network on the AudioCaps dataset \cite{kim2019audiocaps} in order to transfer knowledge in both audio and language modalities. Both transfer learning strategies are adapted and compared in the proposed system.

Another problem is the mismatch between the evaluation metrics and the loss function used for text generation. The evaluation metrics are discrete and non-differentiable, thus cannot be optimized directly by back-propagation. Previous works use reinforcement learning by incorporating the evaluation metrics into the optimisation of the learning system \cite{xu2020crnn, liu2017improved}. We analyze the effects of reinforcement learning on audio captioning system. The results show that even though reinforcement learning can improve the score of evaluation metrics, it may impact adversely on the quality of the generated captions, in the sense that some redundant words are introduced in the captions generated. This finding indicates that existing metrics used for caption evaluation do not correlate well with human judgment. Our resulting system\footnote{\url{https://github.com/XinhaoMei/DCASE2021_task6_v2.git}} was ranked in the 3rd place in DCASE 2021 Task 6 and was the highest scoring system without using an ensemble technique.

\begin{figure}[!t]
    \centering
    \includegraphics[width=\columnwidth]{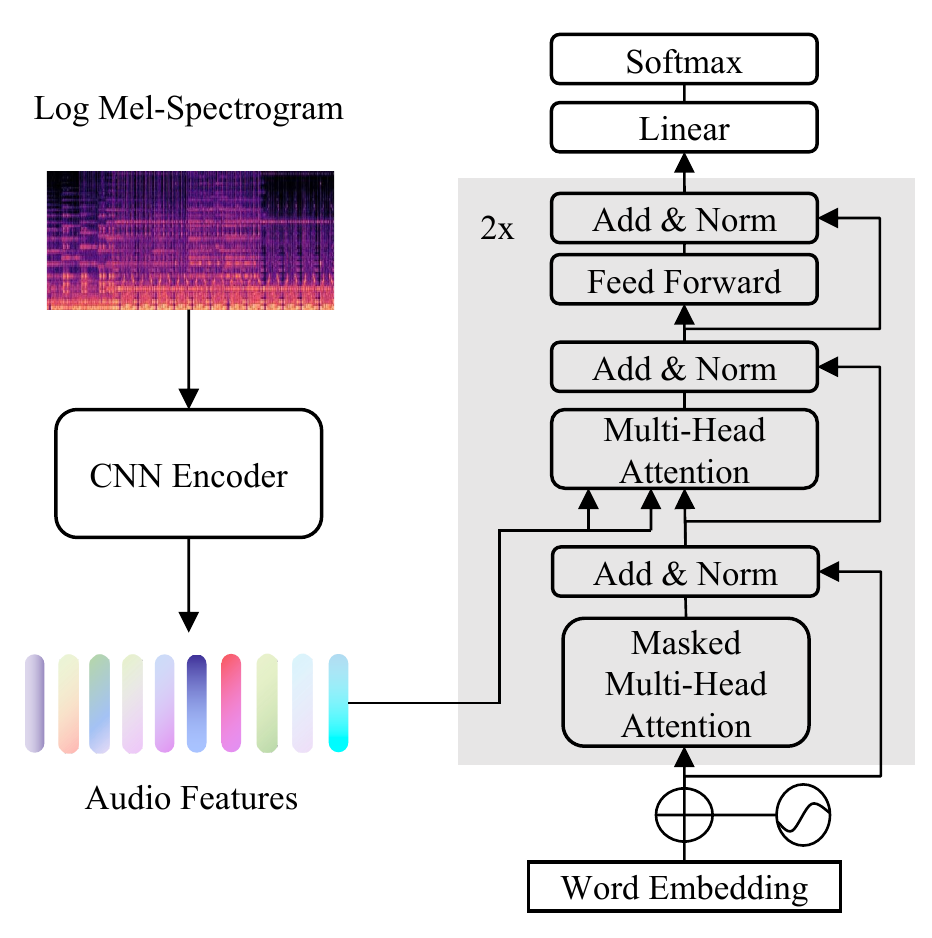}
    \caption{Architecture of the proposed model.}
    \label{fig:network}
\end{figure}

The remaining sections of this paper are organised as follows. In Section \ref{sec:methods}, the proposed model and methods are described in detail. Experimental setup is described in Section \ref{sec:experiments}. Results are shown in Section \ref{sec:results}. Finally, we conclude our work in Section \ref{sec:conclusion}.

\section{Proposed method}
\label{sec:methods}

The proposed model consists of a CNN encoder and a Transformer decoder. The encoder takes the log mel-spectrogram $X \in \mathbb R^{T\times D}$ of an audio clip as input and produces audio feature vector $v \in \mathbb R^{T' \times D'}$. The decoder predicts the posterior probability of the $n$-th word $w_n$ based on the feature vector $v$ produced by the encoder and previously generated words $w_0$ to $w_{n-1}$. Mathematically,
\begin{equation}
  \label{eqn:encoder}
  v = {\rm Enc}(X)
\end{equation}
\begin{equation}
  \label{eqn:decoder}
  p(w_n|v,w_0,...,w_{n-1}) = {\rm Dec}(v, w_0,...,w_{n-1})
\end{equation}

The diagram of the proposed system is shown in Fig.~\ref{fig:network}. 

\subsection{Model architecture}
\subsubsection{Encoder}
\label{sssec:cnn_encoder}

Convolutional neural networks (CNNs) have been used widely in audio processing related works and have shown powerful ability in extracting audio features. A relatively simple 10-layer CNN proposed in the pre-trained audio neural networks (PANNs) \cite{kong2020panns} is used as the encoder to mitigate the over-fitting problem. The 10-layer CNN consists of four convolutional blocks where each has two convolutional layers with a kernel size of $3 \times 3$. The number of channels in each block is \num{64}, \num{128}, \num{256} and \num{512} respectively, and an average pooling layer with a kernel size of $2 \times 2$ is applied between them for down-sampling. After each convolutional layer, batch normalization and ReLU nonlinearity are used. Global average pooling is applied along the frequency axis after the last convolutional block and two fully connected layers are followed to further increase the representation ability and to ensure the dimension of the output is compatible with that of the decoder. 

\subsubsection{Decoder}
\label{sssec:decoder}

The decoder consists of three parts, a word embedding layer, a standard Transformer decoder and a linear layer. Each input word is firstly encoded to a vector of fixed dimension through the word embedding layer. The word embedding layer can be regarded as an embedding look-up matrix of size $V\times d$, where $V$ is the size of the vocabulary and $d$ is the dimension of the word vector. This layer is randomly initialized and kept frozen during the training stage.

Transformer is designed to handle sequential data and shows state-of-the-art performance in generation tasks in the area of natural language processing \cite{vaswani2017attention}. The Transformer decoder is used as the multi-modal decoder here. Word embeddings from the word embedding layer together with the audio features obtained from the encoder are passed to the transformer decoder and are incorporated through a multi-head attention mechanism. As the captions in the datasets are mostly short in length, the decoder used only consists of two transformer decoder blocks with four heads. The dimension of the hidden layer is \num{128}. A linear layer is used at last to output a probability distribution along the vocabulary.

\subsection{Transfer learning}
\label{ssec:transfer}
Transfer learning aims to transfer knowledge from the source domain to the target domain in order to solve the problem caused by insufficient training data and improve the generalization ability of the model. Transfer learning is mostly used in tasks involving single modality. For this cross-modal (i.e. audio to text) translation task, we introduce two transfer learning methods, where the first is transferring from an upstream task while the second is from an in-domain dataset. 

The encoder extracts audio features from an audio clip and is a relatively separate component in the whole model, thus pre-trained audio models can be adapted as the encoder. Different pre-trained audio models have recently been published which can extract generalized audio patterns. PANNs \cite{kong2020panns} are the models pre-trained on the AudioSet dataset for an audio tagging task which have achieved state-of-the-art performance in many downstream audio pattern recognition tasks. One of the PANNs is used to initialize the parameters in the encoder in order to overcome the data scarcity problem and extract generalized audio features. 

There are many powerful pre-trained language models for text generation tasks \cite{xu2021pretrained}. However, since pre-trained language models do not have encoder-decoder attention modules, it is not feasible to directly use a pre-trained language model as the cross-modal decoder here. In order to to transfer knowledge in both modalities, AudioCaps, the largest audio captioning dataset currently available, is introduced to pre-train the proposed model, which allows transfer learning to be applied in both audio and language modalities.


\subsection{Reinforcement learning}
\label{ssec:reinforce}
The training objective of audio captioning systems is usually to optimize the cross-entropy (CE) loss. That is, the model parameters $\theta$ are trained to minimize
\begin{equation}
  \label{eqn:ce_loss}
  \mathcal L_{\rm CE}(\theta)= -\frac{1}{T} \sum_{t=1}^T \log{p(y_t|y_{1:t-1},v, \theta)}
\end{equation}
where $y_t$ is the ground-truth word at time step $t$.
The model is trained with ``teacher forcing'' strategy, i.e. each word to be predicted is conditioned on previous ground-truth words in the training stage, while it is conditioned on previous output words in the test stage. This discrepancy leads to error accumulation during text generation in the test stage and is known as ``exposure bias'' \cite{rennie2017self}. Another problem is the mismatch between the training objective and the evaluation metrics. The performance of captioning systems is evaluated by discrete metrics, which are non-differentiable and cannot be directly optimized by back-propagation. To address these two problems, reinforcement learning with the policy gradient (PG) method is used to optimize the evaluation metrics considered and to directly improve the scores in terms of these metrics \cite{xu2020crnn, rennie2017self, liu2017improved}.

Reinforcement learning makes it possible to directly back-propagate the evaluation metrics in the form of a reward, which is computed by an evaluation metric. In our work, the model is trained to minimize the the negative expected reward:
\begin{equation}
  \label{eqn:rl_loss}
  \mathcal L_{\rm RL}(\theta)= -\mathbb E_{w^s \sim p_\theta}[r(w^s)],
\end{equation}
where $w^s=(s_1^s,...,w_T^s)$ and $w_t^s$ is the word sampled from the model at time step $t$. To compute the gradient of the negative reward, we choose the self-critical sequence training (SCST) method \cite{rennie2017self}, which directly optimizes the true, sequence-level evaluation metric, but avoids learning an estimate of expected future rewards as a baseline. The expected gradient with a single sample $w^s \sim p_\theta$ can be approximated as:
\begin{equation}
  \label{eqn:rl_gradient}
  \nabla_\theta \mathcal L_{\rm RL}(\theta) \approx -(r(w^s) - r(\hat{w}))\nabla_\theta \log p_\theta (w^s),
\end{equation}
where $r(\hat{w})$ is the reward computed by the current model using a greedy inference algorithm.

\section{Experiments}
\label{sec:experiments}

\subsection{Datasets}
\label{ssec:dataset}

\subsubsection{Clotho}
\label{sssec:clotho}

Clotho \cite{drossos2020clotho} is an audio captioning dataset whose audio clips are all collected from the Freesound archive. To encourage caption diversity, each audio clip is provided with five captions annotated by different
annotators. The duration of the audio clips ranges uniformly from \num{15} to \num{30} seconds. 
All the captions contain eight to \num{20} words.

Clotho v2 contains \num{3839} audio clips with \num{19200} captions in the development split, and \num{1045} audio clips with \num{5225} captions in the validation and evaluation split, respectively. We merge the training and validation split together, which gives a new training set with \num{4884} audio clips. 

During training, each audio clip is combined with one of its five captions as a training sample. During evaluation, all five ground-truth captions of an audio clip are used as references and compared with the predicted caption for metric computation.

\subsubsection{AudioCaps}
\label{sssec:audiocaps}
AudioCaps \cite{kim2019audiocaps} is the largest audio captioning dataset currently available, which contains around 50k audio clips sourced from AudioSet with a duration of 10 seconds. AudioCaps is divided into three splits with \num{49274} audio clips in the training set, \num{494} and \num{957} audio clips in the validation and test set, respectively. Each audio clip contains one caption in the training set, while each contains five captions in the validation and test sets. The length of the captions varies, with some containing only three words while some having more than \num{20} words.

\subsection{Data pre-processing}
\label{ssec:data_process}
The input features we used are 64-dimensional log mel-spectrograms obtained using a \num{1024}-point Hanning window with a hop size of \num{512}-points. SpecAugment \cite{park2019specaugment} is used to augment data during training, which operates on the log mel-spectrogram of an audio clip using frequency masking and time masking. 

All captions in the two datasets are transformed to lower case with punctuation removed. Two special tokens ``\texttt{\textless sos\textgreater}'' and ``\texttt{\textless eos\textgreater}'' are padded at the beginning and end of each caption. The vocabulary of the Clotho dataset contains \num{4367} words. As Clotho and AudioCaps have distinct vocabularies, for transfer learning from AudioCaps to Clotho, these two vocabularies are merged together which give a vocabulary containing \num{6636} words. 

\begin{table*}[!t]
\centering
\begin{tabular}[\linewidth]{c c c c c c c c c c} 
 \hline
 Model & BLEU$_{1}$ & BLEU$_{2}$ & BLEU$_{3}$ & BLEU$_{4}$ & ROUGE$_{L}$ & METERO & CIDEr & SPICE & SPIDEr \\ 
 \hline
 Baseline & 0.525 & 0.344 & 0.237 & 0.163 & 0.359 & 0.154 & 0.352 & 0.100 & 0.226 \\
 B+PANNs & 0.564 & 0.375 & 0.255 & 0.171 & 0.383 & 0.172 & 0.421 & 0.120 & 0.270 \\
 B+PANNs+AC & 0.561 & 0.374 & 0.257 & 0.174 & 0.379 & 0.171 & 0.426 & 0.124 & 0.275 \\
 B+PANNs+RL & 0.639 & 0.415 & 0.276 & 0.174 & 0.401 & 0.186 & 0.452 & 0.131 & 0.292 \\
 B+PANNs+AC+RL & 0.634 & \textbf{0.423} & \textbf{0.288} & \textbf{0.185} & \textbf{0.410} & \textbf{0.187} & \textbf{0.476} & \textbf{0.134} & \textbf{0.305} \\
 \hline
 SJTU  \cite{xu2021_sjtu} & 0.643 & - & - & 0.163 & 0.404 & 0.178 & 0.449 & 0.123 & 0.286 \\
 SJTU\_ensemble  \cite{xu2021_sjtu} & \textbf{0.657} & - & - & 0.174 & 0.408 & 0.182 & 0.468 & 0.123 & 0.295 \\
 \hline
\end{tabular}

\caption{Scores of our models on the Clotho v2 evaluation set. Baseline (B): the proposed model trained from scratch. RL: the model fine-tuned using reinforcement learning. PANNs: use PANNs as the audio encoder. AC: the whole model pre-trained on the AudioCaps dataset. Higher score indicates better system performance.}
\label{table:tab_results} 
\end{table*}

\begin{table*}[ht]
\centering
\begin{tabular}[\linewidth]{c | c | c} 
 \hline
  Examples & B+PANNs (w/o RL) & B+PANNS+RL (w/ RL) \\ 
 \hline
 example 1 & a crowd of people are talking and cheering & a crowd of people are talking and in the background \\
 example 2 & a car is driving down the road with the windows open & a car is driving by on and then the engine of a vehicle  \\
 example 3 & someone is playing a guitar with a stick & a guitar is being played on a guitar in the background \\
 example 4 & a machine is running at a constant rate & a machine is running and a in the background \\
 example 5 & a police car with a siren blaring in the background & a siren is blaring while sirens are blaring in the background \\
 \hline
\end{tabular}

\caption{Examples of selected captions generated by the model ``B+PANNs'' and ``B+PANNs+RL''. }
\label{table:tab_caps} 
\end{table*}

\subsection{Experimental setups}
\label{ssec:exp_setup}
The whole model is trained using Adam \cite{kingma2014adam} optimizer with a batch size of \num{32}. Warm-up is used in the first \num{5} epochs to linearly increase the learning rate to the initial learning rate. The learning rate is then decreased to \num{1/10} of itself every \num{10} epochs. Dropout with rate \num{0.2} is applied in the proposed model to mitigate the over-fitting problem. To improve the generalization ability of the model and avoid over-confident prediction, label smoothing \cite{szegedy2016labelsmoothing} with $\epsilon=0.1$ is used in all our experiments. During the inference stage, beam search with a beam size up to \num{5} is used to improve the decoding performance. 

For cross-entropy training, the model is directly trained on the Clotho dataset for \num{30} epochs or firstly pre-trained on the AudioCaps dataset for \num{30} epochs then fine-tuned on Clotho dataset for \num{30} epochs with an initial learning rate of \num{1e-3}. The best model in terms of the SPIDE$_r$ score is selected to optimize CIDER$_r$ score using reinforcement learning for \num{60} epochs with a constant learning rate of \num{5e-5} (in the DCASE challenge, we ran reinforcement learning for \num{25} epochs with a constant learning rate of \num{1e-4} \cite{xinhao2021_t6}).

\subsection{Evaluation metrics}
\label{ssec:metrics}
In the DCASE 2021 Task 6, audio captioning systems are evaluated by machine translation metrics (BLEU$_{n}$, ROUGE$_{l}$ and METEOR) and captioning metrics (CIDE$_{r}$, SPICE and SPIDE$_{r}$). BLEU$_{n}$ \cite{papineni2002bleu} is calculated as a weighted geometric mean of modified precision of n-grams. ROUGE$_{l}$ \cite{lin2004rouge} calculates F-measures based on the longest common subsequence. METEOR \cite{lavie2007meteor} measures a harmonic mean of precision and recall based on word level matches between the candidate sentence and references. CIDE$_{r}$ \cite{vedantam2015cider} applies term frequency inverse document frequency (TF-IDF) weights to n-grams and calculates the cosine similarity between them. SPICE \cite{anderson2016spice} transforms captions into scene graphs and calculates F-score based on tuples in them. SPIDE${r}$ \cite{liu2017improved} is a linear combination of SPICE and CIDE$_{r}$, the SPICE score ensures captions are semantically faithful to the audio clip, while CIDE$_{r}$ score ensures captions are syntactically fluent. 

\begin{table}[!ht]
    \centering
    \begin{tabular}{c | c}
    \hline 
   Model & \# audio clips \\
    \hline 
    B+PANNs (w/o RL) & 155 \\
    B+PANNS+RL (w/ RL) &  765 \\
    Ground-truth & 302 \\
    \hline
    \end{tabular}
    \caption{The number of audio clips containing ``in the background'' in generated and ground-truth captions in the evaluation set.}
    \label{table:cap_statics}
\end{table}
\section{Results}
\label{sec:results}
Table \ref{table:tab_results} presents the performances of the proposed system on the Clotho v2 evaluation set. The proposed system is compared with SJTU's system \cite{xu2021_sjtu} which won second place in DCASE 2021 Task 6 and shows state-of-the-art performance in audio captioning \cite{xu2021investigating}. SJTU's system is based on a ``CNN+RNN'' architecture,  transfer learning and reinforcement learning are also used. In addition, an ensemble strategy is adopted in their system to enhance the model performance. As can be seen in Table \ref{table:tab_results}, our best model outperforms SJTU's ensemble model in most evaluation metrics (except BLEU$_1$), which shows the effectiveness of the proposed model. 

Ablation studies are carried out to investigate the effects of each proposed component. From the experimental results, both transfer learning and reinforcement learning can improve system performance with respect to all the evaluation metrics. For transfer learning, the pre-trained audio encoder (PANNs) significantly improve all the metrics as compared to the system trained from scratch, which indicates that a powerful audio encoder is rather important in this cross-modal translation task. Pre-training on the AudioCaps dataset slightly improves most metrics, which confirms that transfer learning in both audio and language modalities performs better than that in a single modality only. 

Reinforcement learning also improves all the evaluation metrics, although it is only used to optimize CIDE$_r$ score. However, it is somewhat surprising that reinforcement learning may impact adversely on the quality of the generated captions. First, reinforcement learning may lead to captions  syntactically incorrect, introduces some repetitive words and generates incomplete captions. As shown in Table \ref{table:tab_caps}, we present five example captions generated by model ``B+PANNs'' and ``B+PANNs+RL'' to demonstrate this observation. Second, after the optimization with reinforcement learning, most captions are appended a phrase ``in the background'' which was not in their ground truth captions. Table \ref{table:cap_statics} shows the statistics of the number of audio clips for which the generated captions contain ``in the background'' before and after using reinforcement learning. There are \num{302} audio clips whose ground-truth captions contain ''in the background''. After using reinforcement learning, \num{765} predicted captions contain ``in the background'', five times more than those without the use of reinforcement learning. These findings suggest that the existing evaluation metrics may not be able to fully reflect the effectiveness of an audio captioning system, or neither are they consistent with human judgement.

\section{Conclusion}
\label{sec:conclusion}
We have presented a ``CNN+Transformer'' audio captioning system with transfer and reinforcement learning and carried out ablation studies on the proposed methods. The results suggest that transfer and reinforcement learning can both improve the performance in terms of the evaluation metrics, while reinforcement learning may impact adversely on the quality of the generated captions. This finding indicates that the existing evaluation metrics used in the captioning system may not fully reflect the quality of captions. Further research should be carried out to find evaluation metrics which matches well with human judgment. 

\section{ACKNOWLEDGMENT}
\label{sec:ack}
This work is partly supported by grant EP/T019751/1 from the Engineering and Physical Sciences Research Council (EPSRC), a Newton Institutional Links Award from the British Council, titled ``Automated Captioning of Image and Audio for Visually and Hearing Impaired" (Grant number 623805725) and a Research Scholarship from the China Scholarship Council (CSC) No. 202006470010. 

\bibliographystyle{IEEEtranN}
\bibliography{refs}

\begin{thebibliography}{26}
\providecommand{\natexlab}[1]{#1}
\providecommand{\url}[1]{#1}
\csname url@samestyle\endcsname
\providecommand{\newblock}{\relax}
\providecommand{\bibinfo}[2]{#2}
\providecommand{\BIBentrySTDinterwordspacing}{\spaceskip=0pt\relax}
\providecommand{\BIBentryALTinterwordstretchfactor}{4}
\providecommand{\BIBentryALTinterwordspacing}{\spaceskip=\fontdimen2\font plus
\BIBentryALTinterwordstretchfactor\fontdimen3\font minus
  \fontdimen4\font\relax}
\providecommand{\BIBforeignlanguage}[2]{{%
\expandafter\ifx\csname l@#1\endcsname\relax
\typeout{** WARNING: IEEEtranN.bst: No hyphenation pattern has been}%
\typeout{** loaded for the language `#1'. Using the pattern for}%
\typeout{** the default language instead.}%
\else
\language=\csname l@#1\endcsname
\fi
#2}}
\providecommand{\BIBdecl}{\relax}
\BIBdecl

\bibitem[Drossos et~al.(2017)Drossos, Adavanne, and
  Virtanen]{drossos2017automated}
K.~Drossos, S.~Adavanne, and T.~Virtanen, ``Automated audio captioning with
  recurrent neural networks,'' in \emph{IEEE Workshop on Applications of Signal
  Processing to Audio and Acoustics}.\hskip 1em plus 0.5em minus 0.4em\relax
  IEEE, 2017, pp. 374--378.

\bibitem[Kim et~al.(2019)Kim, Kim, Lee, and Kim]{kim2019audiocaps}
C.~D. Kim, B.~Kim, H.~Lee, and G.~Kim, ``Audiocaps: Generating captions for
  audios in the wild,'' in \emph{Proceedings of the 2019 Conference of the
  North American Chapter of the Association for Computational Linguistics:
  Human Language Technologies}, 2019, pp. 119--132.

\bibitem[Drossos et~al.(2020)Drossos, Lipping, and Virtanen]{drossos2020clotho}
K.~Drossos, S.~Lipping, and T.~Virtanen, ``Clotho: An audio captioning
  dataset,'' in \emph{IEEE International Conference on Acoustics, Speech and
  Signal Processing (ICASSP)}.\hskip 1em plus 0.5em minus 0.4em\relax IEEE,
  2020, pp. 736--740.

\bibitem[Nguyen et~al.(2020)Nguyen, Drossos, and Virtanen]{nguyen2020temporal}
K.~Nguyen, K.~Drossos, and T.~Virtanen, ``Temporal sub-sampling of audio
  feature sequences for automated audio captioning,'' \emph{arXiv preprint
  arXiv:2007.02676}, 2020.

\bibitem[Chen et~al.(2020)Chen, Wu, Wang, Zhang, Nian, Li, and
  Shao]{chen2020audio}
K.~Chen, Y.~Wu, Z.~Wang, X.~Zhang, F.~Nian, S.~Li, and X.~Shao, ``Audio
  captioning based on transformer and pre-trained cnn,'' in \emph{Proceedings
  of the Detection and Classification of Acoustic Scenes and Events Workshop},
  2020, pp. 21--25.

\bibitem[Tran et~al.(2020)Tran, Drossos, and Virtanen]{tran2020wavetransformer}
A.~Tran, K.~Drossos, and T.~Virtanen, ``Wavetransformer: A novel architecture
  for audio captioning based on learning temporal and time-frequency
  information,'' \emph{arXiv preprint arXiv:2010.11098}, 2020.

\bibitem[Xu et~al.(2020)Xu, Dinkel, Wu, and Yu]{xu2020crnn}
X.~Xu, H.~Dinkel, M.~Wu, and K.~Yu, ``A crnn-gru based reinforcement learning
  approach to audio captioning,'' in \emph{Proceedings of the Detection and
  Classification of Acoustic Scenes and Events Workshop}, 2020, pp. 225--229.

\bibitem[Wang et~al.(2020)Wang, Yang, Zou, and Chong]{wang2020automated}
H.~Wang, B.~Yang, Y.~Zou, and D.~Chong, ``Automated audio captioning with
  temporal attention,'' DCASE2020 Challenge, Tech. Rep., 2020.

\bibitem[Koizumi et~al.(2020)Koizumi, Masumura, Nishida, Yasuda, and
  Saito]{koizumi2020transformer}
Y.~Koizumi, R.~Masumura, K.~Nishida, M.~Yasuda, and S.~Saito, ``A
  transformer-based audio captioning model with keyword estimation,''
  \emph{arXiv preprint arXiv:2007.00222}, 2020.

\bibitem[Takeuchi et~al.(2020)Takeuchi, Koizumi, Ohishi, Harada, and
  Kashino]{takeuchi2020effects}
D.~Takeuchi, Y.~Koizumi, Y.~Ohishi, N.~Harada, and K.~Kashino, ``Effects of
  word-frequency based pre-and post-processings for audio captioning,''
  \emph{arXiv preprint arXiv:2009.11436}, 2020.

\bibitem[Xu et~al.(2021{\natexlab{a}})Xu, Dinkel, Wu, Xie, and
  Yu]{xu2021investigating}
X.~Xu, H.~Dinkel, M.~Wu, Z.~Xie, and K.~Yu, ``Investigating local and global
  information for automated audio captioning with transfer learning,'' in
  \emph{IEEE International Conference on Acoustics, Speech and Signal
  Processing (ICASSP)}.\hskip 1em plus 0.5em minus 0.4em\relax IEEE, 2021, pp.
  905--909.

\bibitem[Liu et~al.(2017)Liu, Zhu, Ye, Guadarrama, and Murphy]{liu2017improved}
S.~Liu, Z.~Zhu, N.~Ye, S.~Guadarrama, and K.~Murphy, ``Improved image
  captioning via policy gradient optimization of spider,'' in \emph{Proceedings
  of the IEEE international conference on Computer Vision}, 2017, pp. 873--881.

\bibitem[Kong et~al.(2020)Kong, Cao, Iqbal, Wang, Wang, and
  Plumbley]{kong2020panns}
Q.~Kong, Y.~Cao, T.~Iqbal, Y.~Wang, W.~Wang, and M.~D. Plumbley, ``Panns:
  Large-scale pretrained audio neural networks for audio pattern recognition,''
  \emph{IEEE/ACM Transactions on Audio, Speech, and Language Processing},
  vol.~28, pp. 2880--2894, 2020.

\bibitem[Vaswani et~al.(2017)Vaswani, Shazeer, Parmar, Uszkoreit, Jones, Gomez,
  Kaiser, and Polosukhin]{vaswani2017attention}
A.~Vaswani, N.~Shazeer, N.~Parmar, J.~Uszkoreit, L.~Jones, A.~N. Gomez,
  {\L}.~Kaiser, and I.~Polosukhin, ``Attention is all you need,'' in
  \emph{Advances in Neural Information Processing Systems}, 2017, pp.
  5998--6008.

\bibitem[Xu et~al.(2021{\natexlab{b}})Xu, Zhengyan, Ning, Yuxian, Xiao, Yuqi,
  Jiezhong, Liang, Wentao, Minlie, Qin, Yanyan, Yang, Zhiyuan, Zhiwu, Xipeng,
  Ruihua, Jie, Ji-Rong, Jinhui, Xin, and Jun]{xu2021pretrained}
H.~Xu, Z.~Zhengyan, D.~Ning, G.~Yuxian, L.~Xiao, H.~Yuqi, Q.~Jiezhong,
  Z.~Liang, H.~Wentao, H.~Minlie, J.~Qin, L.~Yanyan, L.~Yang, L.~Zhiyuan,
  L.~Zhiwu, Q.~Xipeng, S.~Ruihua, T.~Jie, W.~Ji-Rong, Y.~Jinhui, Z.~W. Xin, and
  Z.~Jun, ``Pre-trained models: Past, present and future,'' 2021.

\bibitem[Rennie et~al.(2017)Rennie, Marcheret, Mroueh, Ross, and
  Goel]{rennie2017self}
S.~J. Rennie, E.~Marcheret, Y.~Mroueh, J.~Ross, and V.~Goel, ``Self-critical
  sequence training for image captioning,'' in \emph{Proceedings of the IEEE
  conference on Computer Vision and Pattern Recognition}, 2017, pp. 7008--7024.

\bibitem[Park et~al.(2019)Park, Chan, Zhang, Chiu, Zoph, Cubuk, and
  Le]{park2019specaugment}
D.~S. Park, W.~Chan, Y.~Zhang, C.-C. Chiu, B.~Zoph, E.~D. Cubuk, and Q.~V. Le,
  ``Specaugment: A simple data augmentation method for automatic speech
  recognition,'' \emph{arXiv preprint arXiv:1904.08779}, 2019.

\bibitem[Xu et~al.(2021{\natexlab{c}})Xu, Xie, Wu, and Yu]{xu2021_sjtu}
X.~Xu, Z.~Xie, M.~Wu, and K.~Yu, ``The {SJTU} system for {DCASE2021} challenge
  task 6: Audio captioning based on encoder pre-training and reinforcement
  learning,'' DCASE2021 Challenge, Tech. Rep., July 2021.

\bibitem[Kingma and Ba(2014)]{kingma2014adam}
D.~P. Kingma and J.~Ba, ``Adam: A method for stochastic optimization,''
  \emph{arXiv preprint arXiv:1412.6980}, 2014.

\bibitem[Szegedy et~al.(2016)Szegedy, Vanhoucke, Ioffe, Shlens, and
  Wojna]{szegedy2016labelsmoothing}
C.~Szegedy, V.~Vanhoucke, S.~Ioffe, J.~Shlens, and Z.~Wojna, ``Rethinking the
  inception architecture for computer vision,'' in \emph{Proceedings of the
  IEEE Conference on Computer Vision and Pattern Recognition}, 2016, pp.
  2818--2826.

\bibitem[Mei et~al.(2021)Mei, Huang, Liu, Chen, Wu, Wu, Zhao, Li, Ko, Tang,
  Shao, Plumbley, and Wang]{xinhao2021_t6}
X.~Mei, Q.~Huang, X.~Liu, G.~Chen, J.~Wu, Y.~Wu, J.~Zhao, S.~Li, T.~Ko, H.~L.
  Tang, X.~Shao, M.~D. Plumbley, and W.~Wang, ``An encoder-decoder based audio
  captioning system with transfer and reinforcement learning for {DCASE}
  challenge 2021 task 6,'' DCASE2021 Challenge, Tech. Rep., July 2021.

\bibitem[Papineni et~al.(2002)Papineni, Roukos, Ward, and
  Zhu]{papineni2002bleu}
K.~Papineni, S.~Roukos, T.~Ward, and W.-J. Zhu, ``Bleu: a method for automatic
  evaluation of machine translation,'' in \emph{Proceedings of the 40th annual
  meeting of the Association for Computational Linguistics}, 2002, pp.
  311--318.

\bibitem[Lin(2004)]{lin2004rouge}
C.-Y. Lin, ``Rouge: A package for automatic evaluation of summaries,'' in
  \emph{Text summarization branches out}, 2004, pp. 74--81.

\bibitem[Lavie and Agarwal(2007)]{lavie2007meteor}
A.~Lavie and A.~Agarwal, ``Meteor: An automatic metric for mt evaluation with
  high levels of correlation with human judgments,'' in \emph{Proceedings of
  the Second Workshop on Statistical Machine Translation}, 2007, pp. 228--231.

\bibitem[Vedantam et~al.(2015)Vedantam, Lawrence~Zitnick, and
  Parikh]{vedantam2015cider}
R.~Vedantam, C.~Lawrence~Zitnick, and D.~Parikh, ``Cider: Consensus-based image
  description evaluation,'' in \emph{Proceedings of the IEEE conference on
  Computer Vision and Pattern Recognition}, 2015, pp. 4566--4575.

\bibitem[Anderson et~al.(2016)Anderson, Fernando, Johnson, and
  Gould]{anderson2016spice}
P.~Anderson, B.~Fernando, M.~Johnson, and S.~Gould, ``Spice: Semantic
  propositional image caption evaluation,'' in \emph{European Conference on
  Computer Vision}.\hskip 1em plus 0.5em minus 0.4em\relax Springer, 2016, pp.
  382--398.

\end{thebibliography}

\end{sloppy}
\end{document}